\documentclass[twocolumn,english,aps,prl,showpacs]{revtex4}
\usepackage[T1]{fontenc}
\usepackage{amsmath,graphicx,amssymb,epsfig,babel,dsfont}

\renewcommand{\Im}{{\rm Im}}

\newcommand{\rD}{{\rm D}}
\newcommand{\rS}{{\rm S}}
\newcommand{\rG}{{\rm G}}
\newcommand{\rE}{{\rm E}}
\newcommand{\rF}{{\rm F}}
\newcommand{\rs}{{\rm s}}
\newcommand{\rp}{{\rm p}}
\newcommand{\rc}{{\rm c}}
\newcommand{\kB}{k_{\rm B}}

\begin{document}

\title{Near-field thermal transistor}

\author{Philippe Ben-Abdallah}
\email{pba@institutoptique.fr}
\affiliation{Laboratoire Charles Fabry,UMR 8501, Institut d'Optique, CNRS, Universit\'{e} Paris-Sud 11,
2, Avenue Augustin Fresnel, 91127 Palaiseau Cedex, France.}
\author{Svend-Age Biehs}
\email{s.age.biehs@uni-oldenburg.de}
\affiliation{Institut f\"{u}r Physik, Carl von Ossietzky Universit\"{a}t, D-26111 Oldenburg, Germany.}

\date{\today}


\begin{abstract}
Using a block of three separated solid elements, a thermal source and drain together with a gate made of an insulator-metal transition 
material exchanging near-field thermal radiation, we introduce a nanoscale analog of a field-effect transistor which is able to control 
the flow of heat exchanged by evanescent thermal photons between two bodies. By changing the gate temperature around its critical value, the heat 
flux exchanged between the hot body (source) and the cold body (drain) can be reversibly switched, amplified, and modulated by a tiny action 
on the gate. Such a device could find important applications in the domain of nanoscale thermal management and it opens up new perspectives 
concerning the development of contactless thermal circuits intended for information processing using the photon current rather than the 
electric current.  
\end{abstract}

\maketitle

\begin{figure}[Hhbt]
\includegraphics[scale=0.3]{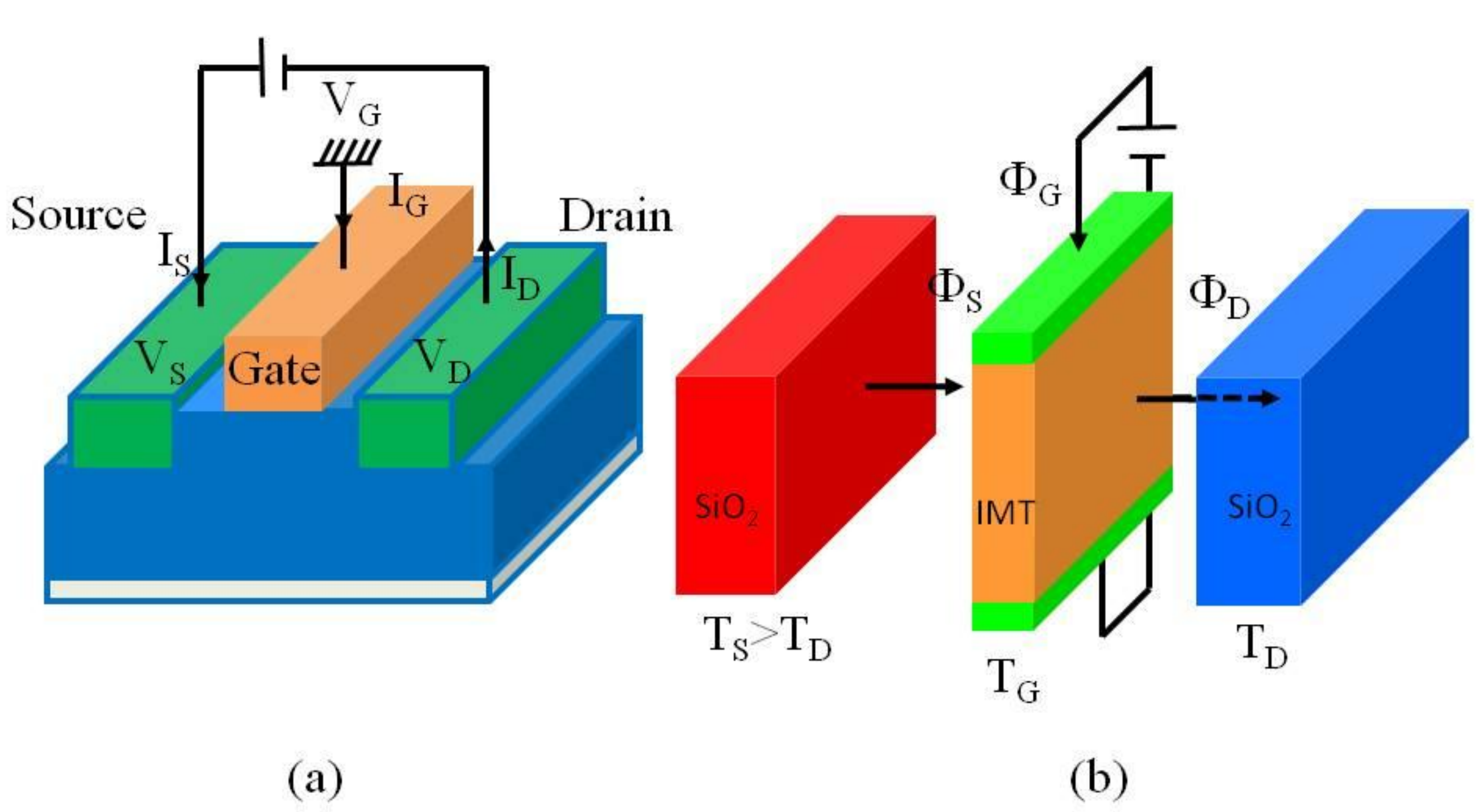}
\caption{ $\bold{Electric}$ $\bold{and}$ $\bold{radiative}$ $\bold{thermal}$ $\bold{transistor}$. Classical field-effect transistor (a): a three-terminal device known as source, gate and drain, which correspond to the emitter, base, and collector of electrons. The gate is used to actively 
control, by applying a potential on it, the apparent conductivity of the channel between the source and the drain. Radiative thermal transistor (b):  
a layer of IMT material (the gate) is placed at subwavelength distances from two thermal reservoirs (the source and the drain). The temperature $T_\rS$ 
and $T_\rD$ are fixed so that $T_\rS>T_\rD$. The temperature $T_\rG$ of the gate can be modulated from the external environment around its steady-state 
temperature $T^{\rm eq}_\rG$ (which corresponds to the situation where the net flux $\Phi_\rG$ received by the gate vanishes) so that the flux $\Phi_\rD$ received 
by the drain and lost by the source $\Phi_\rS$ can be tuned. When $T^{\rm eq}_\rG$ is a little bit smaller than the critical temperature $T_\rc$ of IMT material 
a small amount of heat applied on the gate induces a strong switching of heat fluxes $\Phi_\rD$ and $\Phi_\rS$ owing to its metal-insulator phase transition.  
\label{Transistor}}
\end{figure}

The electronic solid-state transistor (Fig.~\ref{Transistor}) introduced by Bardeen and Brattain in 1948~\cite{Bardeen} is undoubtedly 
the corner stone of almost all modern systems of information treatment. The classical field effect transistor (FET) which is composed 
by three basic elements, the drain, the source, and the gate, is basically used to control the flux of electrons (the current)  exchanged 
in the channel between the drain and the source by changing the voltage applied on the gate. The physical diameter of this channel is 
fixed, but its effective electrical diameter can be varied by the application of a voltage on the gate. A small change in this voltage 
can cause a large variation in the current from the source to the drain. In 2006 Li et al.~\cite{Casati1} have proposed a thermal 
counterpart of FET by replacing both the electric potentials and the electric currents by thermostats at a fixed temperature and heat 
fluxes carried by phonons through solid segments. Later, several prototypes of phononic thermal logic gates~\cite{BaowenLi2} as well 
as thermal memories (see~\cite{BaowenLiEtAl2012} and Refs. therein) have been developed in order to process information by phonon heat flux 
rather than by electric currents. However, this technology suffers from some weakness of fundamental nature which intrinsically limits its 
performance. One of the main limitations comes probably from the speed of acoustic phonons (heat carriers) which is four or five orders 
of magnitude smaller than the speed of photons. This explains, in part, why so many efforts have been deployed, during the last decades, 
to attempt to develop full optical or at least opto-electronic arcitectures for processing and managing information.

We introduce here a thermal transistor (Fig.~\ref{Transistor}) based on the heat transport by evanescent photons rather than by acoustic waves or electrons. 
This near-field thermal transistor (NFTT) basically consists of a gate made of an insulator-metal transition (IMT) material which is able to qualitatively and 
quantitatively change its optical properties through a small change of its temperature around a critical temperature $T_\rc$. Vanadium dioxide (VO$_2$) is one of 
such materials which undergoes a first-order transition (Mott transition~\cite{Mott}) from a high-temperature metallic phase to a low-temperature insulating 
phase~\cite{Baker} close to room-temperature ($T_\rc=340\,{\rm K}$). Different works have already shown~\cite{van Zwol1,van Zwol2,vanZwol3} that the heat-flux exchanged at 
close separation distances (i.e.\ in the near-field regime) between an IMT material and another medium, can be modulated by several orders of magnitude across 
the phase transition of IMT materials. Further radiative thermal diodes have been recently conceived allowing for rectification~\cite{Starr1936,RobertsWalker2011} 
of heat flux using materials with thermally dependend refractive indices~\cite{OteyEtAl2010,Fan,NefzaouiEtAl2013} and IMT materials~\cite{PBA_APL}. But so far 
controlling the heat flow with contactless systems to get the same functionalities as a classical FET has remained a challenging problem. Here we show that IMT 
materials are very promising candidates for designing efficient gates to control (switch, modulate or amplify) the heat flux exchanged between two media.

To start, let us consider the system as illustrated in Fig.~\ref{Transistor}(b), where two media which we call by analogy with a FET the source and the drain 
labeled by the indices $\rS$ and $\rD$ are maintained at temperatures $T_\rS$ and $T_\rD$ with $T_\rS>T_\rD$ by some thermostats. A thin layer of IMT material 
labeled by $\rG$ having a thickness $\delta$ is placed between both media at a distance $d$ from the source and the drain. Without external excitation, the 
system reaches its steady state for which the net flux $\Phi_\rG$ received by the intermediate medium, the gate, is zero. In this case its temperature $T_\rG$ is 
set by the temperature of the surrounding media, i.e.\ the drain and the source. When a certain amount of heat is added to or removed from the gate (for example by applying 
a voltage difference through a couple of electrodes as illustrated in Fig.~\ref{Transistor} or extracted from it using Peltier elements), its temperature can 
be either increased or reduced around its equilibrium temperature $T^{\rm eq}_\rG$. Hence, the heat flux $\Phi_\rD$ received by the drain and the flux $\Phi_\rS$ lost by the 
source can be tailored accordingly. These heat fluxes correspond to the flux of Poynting vector across any plane separating the gate and the drain 
(and the source and the gate). In a three body system the flux received by photon tunneling by the drain reads~\cite{Messina}
\begin{equation}
  \Phi_{\rD}= \int_0^\infty\!\frac{d\omega}{2\pi}\,\phi_\rD(\omega,d), 
\label{Eq:Flux_D}
\end{equation}
where the monochromatic heat flux is given by
\begin{equation}
\begin{split}
  \phi_{\rD} &= \hbar\omega\sum_{j = \{\rm s,p\}}\int\! \frac{{\rm d}^2 \boldsymbol{\kappa}}{(2 \pi)^2} \, [n_{\rS\rG}(\omega)\mathcal{T}^{\rS/\rG}_j(\omega,\boldsymbol{\kappa}; d)\\
                &+n_{\rG\rD}(\omega)\mathcal{T}^{\rG/\rD}_j(\omega,\boldsymbol{\kappa}; d)].
\end{split}
\label{Eq:Flux_D_1}
\end{equation}
Here $\mathcal{T}^{\rS/\rG}_j$ and $\mathcal{T}^{\rG/\rD}_j$ denote the efficiencies of coupling of each mode $(\omega,\boldsymbol{\kappa})$ between the source and the gate 
and between the gate and the drain for both polarization states $j=\rs,\rp$; $\boldsymbol{\kappa} = (k_x,k_y)^t$ is the wavevector parallel to the surfaces of the multilayer
system. In the above relation $n_{ij}$ denotes the difference of Bose-distributions functions 
$n_i$ and $n_j$ [with $n_{i/j}=(e^{\frac{\hbar \omega}{\kB T_{i/j}}}-1)^{-1}$] at the frequency $\omega$; $\kB$ is Boltzmann's constant and $2 \pi \hbar$ is Planck's constant. 
According to the N-body near-field heat transfer theory presented in Ref.~\cite{Messina}, the transmission coefficients $\mathcal{T}^{\rS/\rG}_j$ and 
$\mathcal{T}^{\rG/\rD}_j$ of the energy carried by each mode written in terms of optical reflection coefficients $\rho_{\rE,j}$ ($\rE = \rS, \rD, \rG$) and transmission 
coefficients $\tau_{\rE,j}$ of each basic element of the system and in terms of reflection coefficients $\rho_{\rE\rF,j}$ of couples of elementary elements~\cite{Messina}
\begin{equation}
\begin{split}
  &\mathcal{T}^{\rS/\rG}_{j}(\omega,\boldsymbol{\kappa},d)\\
  &=\frac{4\mid\tau_{\rG,j}\mid^2 \Im(\rho_{\rS,j})\Im(\rho_{\rD,j})e^{-4\gamma d}}{\mid 1-\rho_{\rS\rG,j}\rho_{\rD,j}e^{-2\gamma d}\mid^2\mid1-\rho_{\rS,j}\rho_{\rG,j}e^{-2\gamma d}\mid^2},\\
  &\mathcal{T}^{\rG/\rD}_{j}(\omega,\boldsymbol{\kappa},d)=\frac{4 \Im(\rho_{\rS\rG,j})\Im(\rho_{\rD,j})e^{-2\gamma d}}{\mid 1-\rho_{\rS\rG,j}\rho_{\rD,j}e^{-2\gamma d}\mid^2}
\end{split}
\label{Trans}
\end{equation}
introducing the imaginary part of the wavevector normal to the surfaces in the multilayer structure $\gamma = \Im(k_z) = \sqrt{\kappa^2 - \omega^2/c^2}$; 
$c$ is the velocity of light in vacuum. Similarly the heat flux lost by the source reads
\begin{equation}
\begin{split}
  \phi_{\rS} &= \hbar\omega\sum_{j = \{\rm s,p\}}\int\! \frac{{\rm d}^2 \boldsymbol{\kappa}}{(2 \pi)^2} \, [n_{\rD\rG}(\omega)\mathcal{T}^{\rD/\rG}_j(\omega,\boldsymbol{\kappa}; d)\\
                &+n_{\rG\rS}(\omega)\mathcal{T}^{\rG/\rS}_j(\omega,\boldsymbol{\kappa}; d)]
\end{split}
\label{Eq:Flux_S_1}
\end{equation}
where the transmission coefficients are analog to those defined in Eq.~(\ref{Trans}) and can be obtained making the substitution $S\leftrightarrow D$.

At steady state, the net heat flux received/emitted by the gate which is just given by the heat flux from the source to the gate minus the heat flux from the gate to the drain vanishes, i.e. $\Phi_\rS = \Phi_\rD$ or 
\begin{equation}
 \Phi_\rG = \Phi_\rS - \Phi_\rD = 0.
\end{equation}
This relation allows us to identify the gate equilibrium temperature $T^{\rm eq}_\rG$ (which is not necessary unique) for given 
temperatures $T_\rS$ and $T_\rD$. Note that out of steady state the heat flux received/emitted by the gate is $\Phi_\rG = \Phi_\rS - \Phi_\rD \neq 0$.
If $\Phi_\rG > 0$ ($\Phi_\rG < 0$) an external flux is added to (removed from) the gate by heating (cooling). 

\begin{figure}[Hhbt]
\includegraphics[scale=0.40]{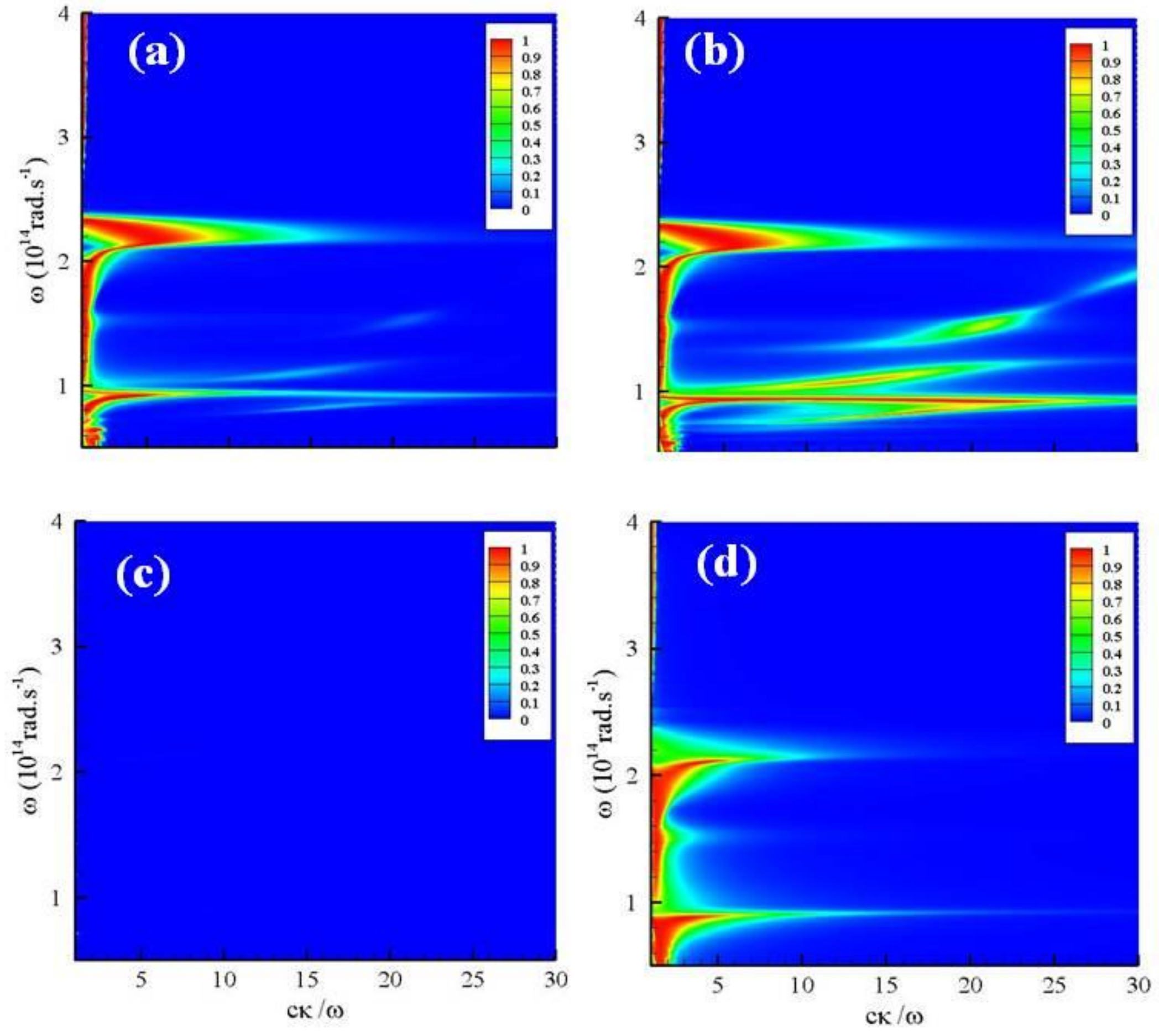}

\caption{$\bold{Transmission}$ $\bold{probabilities}$ $\bold{of}$ $\bold{energy}$ $\bold{carried}$ $\bold{by}$ $\bold{the}$ $\bold{modes}$. 
Efficiency of coupling of modes ($\omega,\boldsymbol{\kappa}$) in a SiO$_2$-VO$_2$-SiO$_2$ system ($\delta=50\,{\rm nm}$  and $d=100\,{\rm nm}$). (a) $\mathcal{T}^{\rS/\rG}_\rp$ and 
(b) $\mathcal{T}^{\rG/\rD}_\rp$ with VO$_2$ in its crystalline state. (c) $\mathcal{T}^{\rS/\rG}_\rp$ and (d) $\mathcal{T}^{\rG/\rD}_\rp$ with VO$_2$ in its amorphous state. 
Wien's frequency (where the transfer is maximum) at $T=340\,{\rm K}$ is $\omega_{\rm Wien}\sim 1.3\times 10^{14}\,{\rm rad/s}$. The dielectric
permittivity of SiO$_2$ is taken from the database~\cite{Palik}.
\label{Transmission}}
\end{figure}

\begin{figure}[Hhbt]
\includegraphics[scale=0.35]{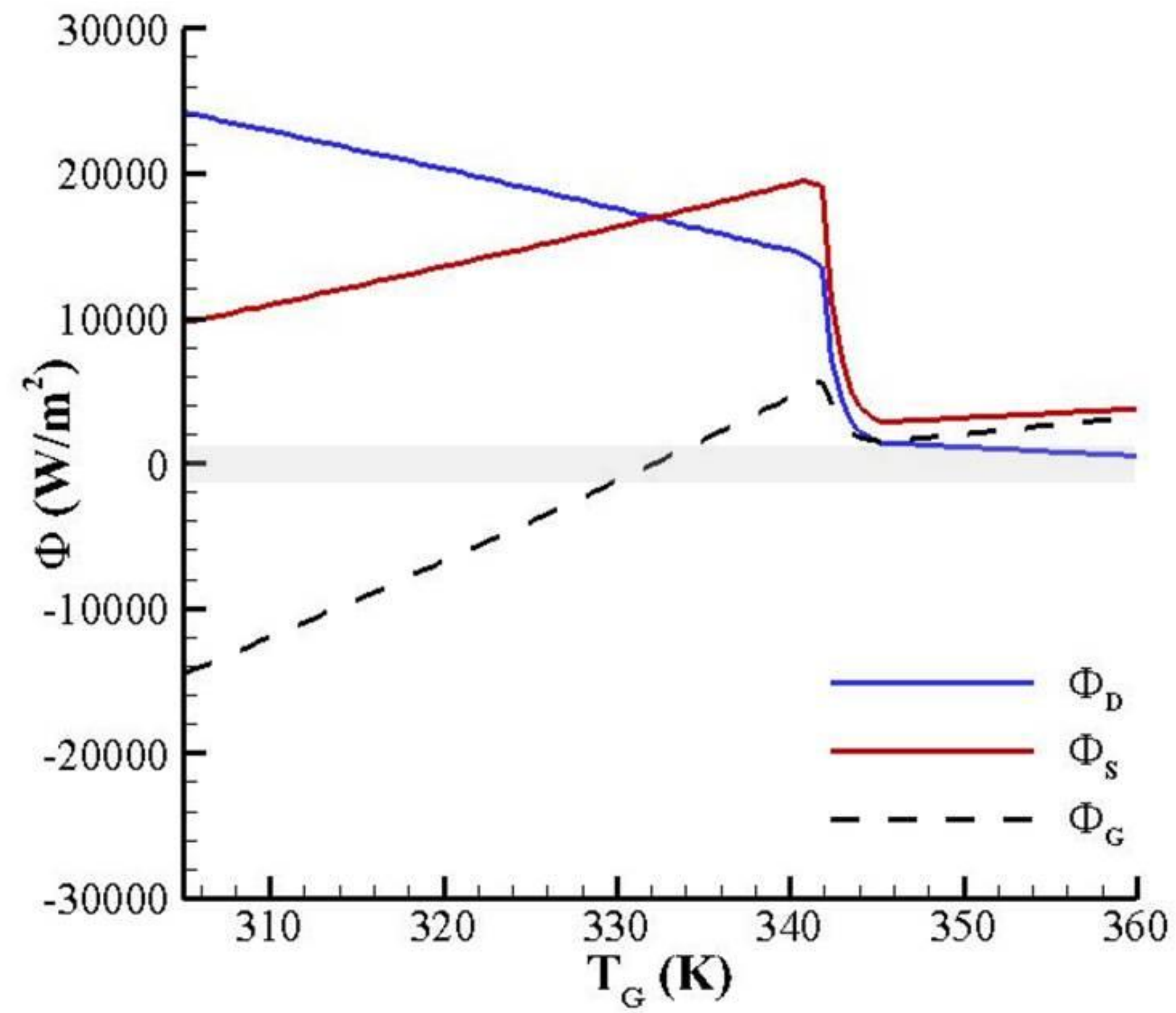}

\caption{$\bold{Operating}$ $\bold{regimes}$ $\bold{of}$ $\bold{near}$$\bold{-}$$\bold{field}$ $\bold{thermal}$ $\bold{transistor}$. When $T^{eq}_G$ is a little 
bit smaller than the critical temperature $T_\rc$ of the IMT material a small amount of heat applied on the gate induces     a strong switching 
of heat fluxes  $\phi_\rD$ and $\phi_\rS$ owing to its phase transition. By changing the flux $\phi_\rG$ supplied to the gate different functions (thermal switching, thermal modulation and 
thermal amplification) can be performed.  The fluxes plotted here correspond to a gate of VO$_2$~\cite{Baker} with thickness $\delta = 50$ nm located at a distance $d = 100$ nm from two massive silica 
samples~\cite{Palik} maintained at $T_\rS=360\,{\rm K}$ and $T_\rD=300\,{\rm K}$.  
\label{Flux}}
\end{figure}

\begin{figure}[Hhbt]
\includegraphics[scale=0.35]{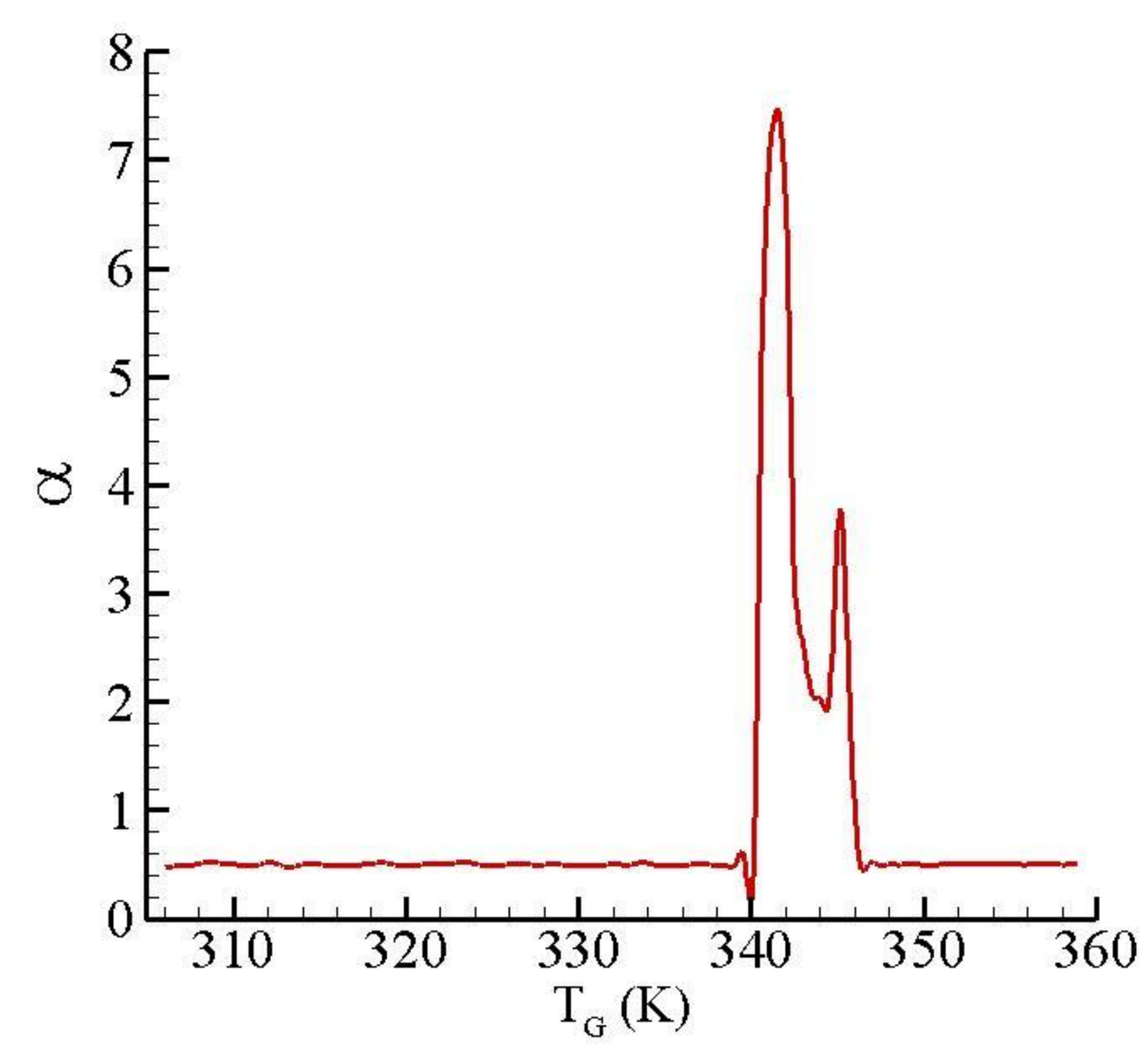}

\caption{$\bold{Amplification}$ $\bold{factor}$ $\bold{\alpha}$ $\bold{of}$ $\bold{the}$ $\bold{NFTT}$. When $T_G\ll T_c$ or $T_G\gg T_c$ the slopes of $\Phi_\rS$ and $\Phi_\rD$ are almost identical (modulo the sign) so that $\alpha\approx 1/2$.  On the contrary, in the close neighborhood of $T_c$ we have $\alpha>1$ owing to the negative differential thermal resistance in the transition region. Here, the parameters of the NFTT are the same as in Fig.~3.  
\label{Amplification}}
\end{figure}

To illustrate the operating modes of NFTT we consider now a system composed by a source and a drain both made of silica (each coupled to a thermostat to 
maintain their local temperatures constant in time) and a gate made of vanadium dioxide VO$_2$. When the gate temperature $T_\rG$ is smaller than its
critical temperature $T_\rc$, then the gate is in its monoclinic phase and it behaves as an uniaxial crystal. On the other hand, when $T_\rG=T_\rc$ the gate 
transits toward its amorphous metallic phase and remains in this state for greater temperatures. We consider here the case where the optical axis 
of VO$_2$ film is orthogonal to its interfaces. The p-polarized transmission coefficients of the energy carried by the modes $(\omega,\boldsymbol{\kappa})$ through such a system 
are plotted in Fig.~\ref{Transmission}. When the gate is in its crystalline state, $\mathcal{T}^{\rS/\rG}_\rp$, which represents the exchange between the source 
and the drain mediated by the presence of the gate [see Fig.~\ref{Transmission}(a)], and $\mathcal{T}^{\rG/\rD}_\rp$, which  corresponds to the exchange between 
the couple source-gate treated as a unique body and the drain [see Fig.~\ref{Transmission}(b)] 
shows an efficient coupling of modes between the different blocks of the system around the resonance frequencies 
$\omega_{\rm SPP1}\sim 1\times 10^{14}\,{\rm rad/s}$ and $\omega_{\rm SPP2}\sim 2\times 10^{14}\,{\rm rad/s}$ of surface waves (surfaces phonon-polaritons) 
supported by both the source and the drain. Below $T_\rc$ all parts of the system support surface waves in the same frequency range close to the thermal
peak frequency $\omega_{\rm Wien}\sim 1.3\times 10^{14}\,{\rm rad/s}$. The anti-crossing curves which appear in Fig.~\ref{Transmission}(a) and (b) result 
from the strong coupling of silica surface phonon-polaritons (SPPs) and the surface waves (symmetric and antisymmetric ones) suppported by the thin VO$_2$ layer. 
Beyond $T_\rc$ the gate becomes amorphous (metallic) and it does not support surface wave anymore. In this case $\mathcal{T}^{\rS/\rG}_\rp$  [see Fig.~\ref{Transmission}(c)] 
vansihes owing to the field screening by the gate. Moreover, as is clearly shown in Fig.~\ref{Transmission}(d),  the coupling  of modes between the couple 
source-gate and the drain at the frequency of surface waves is less efficient for the large parallel values of $\kappa$ reducing so the transfer of heat towards 
the drain, i.e.\ the number of participating modes decreases~\cite{BiehsEtAl2010,JoulainPBA2010}. By using different physical parameters 
(temperatures, sizes, separation distances...) several functions can be assigned to this system which can become either (i) a thermal switch, (ii) a thermal modulator  
or (iii) a thermal amplifier. We discuss below those operating modes. To do so we show Fig.~\ref{Flux} the net heat flux  received by each part of the system 
in a particular configuration where $\delta = 50\, {\rm nm}$, $d = 100\, {\rm nm}$, $T_\rS=360\, {\rm K}$ and $T_\rD = 300\, {\rm K}$. 
\begin{flushleft}
(i)\underline{\textit{Thermal switching :}}
\end{flushleft}
In the situation depicted in Fig.~(\ref{Flux}) we have $T^{\rm eq}_\rG=332\,{\rm K}$. An increase of $T_\rG$ by about 10 degrees ($\Phi_\rG$ is increased by $\sim 10^{-8} W/\mu m^2$) leads, as clearly shown in Fig.~\ref{Flux}, to a reduction of heat flux received by the drain and lost by the source by more than one order of magnitude. That means our
NFTT can be used in two operating modes where $T_\rG$ is slightly below or above the critical temperature $T_\rc$, where in the case $T_\rG < T_\rc$ we are in the 'on' mode
and for $T_\rG > T_\rc$ we are in the 'off' mode.
\begin{flushleft}
(ii)\underline{\textit{Thermal modulation:}}
\end{flushleft}
Over the temperature region around $T^{\rm eq}_\rG$ (gray shadow strip on Fig.~\ref{Flux}) the heat current $\Phi_G$ over the gate remains quite small (i.e. $\Phi_S\sim\Phi_D$) 
while the flux received by the drain or lost by the source can be modulated from high to low values. The thermal inertia of the gate as well as its phase 
transition delay of IMT material define the timescale at which the  modulator can operate. A much larger modulation of fluxes can be achieved with the NFTT 
when using $T^{\rm eq}_\rG \approx T_\rc$. Then the flux can be modulated over one order of magnitude by a small temperature change of $T_\rG$. 
\begin{flushleft}
(iii)\underline{\textit{Thermal amplification:}}
\end{flushleft}
The most important feature of a transistor is its ability to amplify the current or electron flux towards the drain. 
In the region of phase transition around $T_\rc$ we see that an increase of $T_\rG$ leads to a drastic reduction of flux received by the drain. This corresponds to a negative 
differential thermal conductance as recently described for SiC in Ref.~\cite{Fan} (note that this behavior does not violate the second principle of thermodynamics because the heat 
flux continues to flow from the hot to the cold body). Having a negative differential thermal conductance is the key for having an amplification which is defined as (see for example Ref.~\cite{Casati1})
\begin{equation}
  \alpha \equiv \bigl|\frac{\partial \Phi_\rD}{\partial \Phi_\rG}\bigl| =  \frac{1}{\bigl|1 - \frac{\Phi_\rS'}{\Phi_\rD'}\bigr|}
\end{equation}
where
\begin{equation}
  \Phi_{\rS/\rD}' \equiv \frac{\partial \Phi_{\rS/\rD}}{\partial T_G}.
\end{equation}
It can be easily shown that $\alpha = 1/2$ for $T_G$ much smaller or larger than $T_\rc$ where the material properties of VO$_2$ are more or less independent of $T_G$,
since $\Phi_\rS' = - \Phi_\rD'$. On the other hand, inside the transition region of VO$_2$ that means for temperatures around $T_\rc$ the material properties 
of VO$_2$ change drastically showing a negative differential thermal resistance/conductance which leads to an amplification, i.e.\ $\alpha > 1$. This behaviour is qualitatively
demonstrated in Figs.~\ref{Flux} and~\ref{Amplification} where the dielectric permittivity of  VO$_2$ in the transition region is  modelled  using a Bruggeman mixing rule as introduced in Ref.~\cite{Mott}. Hence, by chosing the temperatures such that $T^{\rm eq}_\rG \approx T_\rc$
the NFTT works as an amplifier.

The ability to control the flow of heat at subwavelength scale in complex architectures of solids out of contact, opens up new opportunities for an active thermal 
management for dissipating systems. It also suggests the possibility to develop contactless thermal analogs of electronic devices such as thermal logic gates and thermal 
memories, for processing information by utilizing thermal photons rather than electrons. Unlike other schemes for creating thermal transistors which were so far based  
on the control of acoustic phonons, the present concept authorizes much higher operational speeds (speed of light) and should be very competitive compared to the 
previous ones. We think also that the near-field thermal transistors could find broad applications in MEMS/NEMS technologies and could be used to generate mechanical 
works by modulating the heat flux received by the drain, by using microresonators such as cantilvers in contact with it.

%
%


\end{document}